\documentclass[twocolumn]{article}

\usepackage[a4paper, margin=0.8in]{geometry}
\usepackage{graphicx}
\usepackage{times}
\usepackage{titlesec}
\usepackage{enumitem}
\usepackage{hyperref}
\usepackage{caption}

\usepackage{authblk}
\usepackage[backend=biber]{biblatex}
\title{Composable OS Kernel Architectures for Autonomous Intelligence}
\author[1]{Rajpreet Singh} 
\author[2]{Vidhi Kothari}
\affil[1]{TU Munich, Germany}
\affil[2]{Pace University, New York, USA}
\date{August 2025}

\begin{document}

\maketitle

\begin{abstract}
As intelligent systems permeate edge devices, cloud infrastructure, and embedded real-time environments, this research proposes a new OS kernel architecture for intelligent systems, transforming kernels from static resource managers to adaptive, AI-integrated platforms. Key contributions include: (1) treating Loadable Kernel Modules (LKMs) as AI-oriented computation units for fast sensory and cognitive processing in kernel space; (2) expanding the Linux kernel into an AI-native environment with built-in deep learning inference, floating-point acceleration, and real-time adaptive scheduling for efficient ML workloads; and (3) introducing a Neurosymbolic kernel design leveraging Category Theory and Homotopy Type Theory to unify symbolic reasoning and differentiable logic within OS internals. Together, these approaches enable operating systems to proactively anticipate and adapt to the cognitive needs of autonomous intelligent applications.

\end{abstract}

\section{Introduction}
As the demand for high-performance, adaptive, and intelligent computing continues to grow, traditional operating system (OS) kernels—originally architected for deterministic and general-purpose computing—are proving increasingly inadequate for modern workloads such as AI Agents. Emerging domains such as Agentic Artificial Intelligence (AI) \cite{jia2024agentcentricoperating}, Machine Learning (ML), and the pursuit of Artificial General Intelligence (AGI) necessitate a fundamental rethinking of kernel design and architecture. The classical abstraction layers separating user-space from kernel-space, while historically beneficial for stability and modularity, introduce performance bottlenecks and hinder dynamic adaptability required by contemporary AI workloads.

Modern operating systems need to adapt quickly to new hardware and evolving applications. Rather than rebuilding or restarting the system for each change, modularity at the kernel level provides much-needed flexibility. Loadable Kernel Modules (LKMs) embody this philosophy by allowing the kernel to be extended or modified at runtime, without reboot or recompilation.

Initially designed for integrating device drivers, filesystems, and networking protocols, LKMs now serve a broader, more dynamic role. They can be leveraged as computational building blocks that execute directly within the kernel, enabling rapid, low-latency processing for tasks where every microsecond counts. By operating within kernel space and close to hardware, LKMs reduce the overhead of user-kernel context switches, making them ideal for latency-sensitive operations, including cryptographic routines, real-time data analytics, and even neural network inference. Furthermore, specialized LKMs for tasks like computer vision, audio analysis, and natural language processing can be composed to form high-speed, distributed AI frameworks entirely within the kernel. This enables intelligent edge and embedded systems to execute critical AI workloads with minimal delay.

However, LKMs also introduce unique challenges: their privileged execution context can affect system stability and impose security and memory protection concerns. This calls for a balanced architecture—where only the most latency-critical functions run as kernel modules, while higher-level learning, orchestration, and adaptive decision-making are managed in user space. Such a hybrid approach maximizes both efficiency and safety, paving the way for truly intelligent, responsive operating systems. Despite these advances, the kernel remains fundamentally static, optimized for fixed heuristics and predefined priorities, unable to accommodate the dynamic and probabilistic nature of AI and AGI workloads. As ML models become more complex, involving quintillions of floating-point operations, GPU-accelerated processing, and multi-modal reasoning, there is a growing need for AI-aware kernels—operating systems that integrate intelligent resource scheduling, GPU-native memory management \cite{10.1145/3315508.3329973}, floating-point optimizations, and ML-aware inference mechanisms. These enhancements aim to reduce latency, improve throughput, and enable reproducibility, making kernels more capable of supporting large-scale and real-time ML workloads.

Yet even these AI-aware optimizations fall short when addressing the requirements of AGI—systems that must not only compute but reason, learn, and adapt autonomously. Contemporary kernels lack the capacity to interpret intent, resolve cross-modal dependencies, or perform symbolic reasoning. This necessitates a paradigm shift toward NeuroSymbolic kernel architectures that embed logical reasoning, probabilistic inference, and neural computation directly into the OS substrate. Such kernels would unify symbolic and sub-symbolic processing, enabling high-level, context-aware resource orchestration and decision making. Foundational mathematical frameworks such as Category Theory, Homotopy Type Theory (HoTT), and Linear Logic offer the scaffolding for constructing these modular, composable, and adaptive reasoning systems.

By presenting the vision of co-designing hardware and software to natively support Neurosymbolic reasoning, integrating persistent memory models for knowledge retention, and embedding knowledge graph traversal mechanisms, we lay the groundwork for AGI-native operating systems. These systems will not only optimize performance but also enable emergent intelligence as an intrinsic property of the computing substrate.



\section{Deep in the Linux Kernel}

The "Deep in the Linux Kernel" contribution introduces a pioneering approach to operating system design by integrating deep learning models as Loadable Kernel Modules (LKMs) \cite{websiteKey1} \cite{websiteKey2} \cite{websiteKey3} within the Linux kernel, enabling real-time, low-latency artificial intelligence (AI) processing. This work realizes the vision for AI-aware kernels by leveraging LKMs to minimize user-kernel context switching, optimize hardware utilization, and support multi-modal AI tasks, establishing a foundation for intelligent, high-performance operating systems. As a standalone contribution, it presents a novel hybrid architecture, technical implementation, performance optimizations, a distributed system architecture, and addresses challenges with future directions for advancing artificial general intelligence (AGI).

\subsection{Hybrid Architecture for Real-Time AI Processing}

This contribution proposes a hybrid architecture that integrates kernel-space LKMs for low-latency, modality-specific AI tasks with a user-space orchestrator for high-level cognitive functions. LKMs are dynamically loaded and unloaded using module\_init() and module\_exit() macros, enabling runtime extension of kernel functionality without recompilation or system reboot. The design evolves from a basic LKM that performs simple arithmetic operations (e.g., addition, subtraction, multiplication, division) through a custom system call to a sophisticated tensor-based LKM capable of executing matrix operations essential for neural inference. By operating in kernel space, LKMs leverage direct access to system resources, minimizing overhead for context switching and enabling efficient processing for applications such as edge computing, high-speed networking, and autonomous systems. This aligns with the abstract’s emphasis on real-time AI processing and proximity to hardware, supporting low-latency inference critical for AI/AGI workloads.

\subsection{Technical Implementation of AI-Enabled LKMs}

To demonstrate practical AI integration within the Linux kernel, we begin by implementing a simple Loadable Kernel Module (LKM) \cite{7280137} that performs basic arithmetic operations on two integers. The module is accessible through a custom system call (for example, \texttt{sys\_simple\_compute}, assigned syscall number 548) and supports operations such as addition, subtraction, multiplication, and division, denoted by operation codes (e.g., \texttt{OP\_ADD=0}, \texttt{OP\_SUBTRACT=1}, etc.). The kernel module includes robust error handling, returning standard error codes to user space (such as \texttt{\-EINVAL} for division by zero). Functionality is validated using a user-space application compiled with \texttt{gcc} and dynamically loaded into the kernel with \texttt{insmod}, highlighting the flexibility and ease of deployment provided by LKMs.

Building on this foundation, we can extend our approach to support more advanced computational tasks with a tensor-processing LKM, capable of performing operations on multi-dimensional arrays—fundamental data structures in machine learning. This LKM dynamically allocates memory via \texttt{kmalloc()}, ensuring alignment for efficient vectorized operations (e.g., utilizing AVX\-512 instructions on x86\-64 platforms). To maximize performance, the module employs zero-copy data transfer techniques between user and kernel space, using \texttt{get\_user\_pages()} and \texttt{kmap()}, alongside \texttt{copy\_from\_user()} and \texttt{copy\_to\_user()} for secure and controlled data movement.

The module supports key tensor operations, including element-wise summation and matrix multiplication (\texttt{OP\_SUM}, \texttt{OP\_MATRIX\_MULT}), while performing checks to ensure input array dimensions are valid (e.g., verifying the column count of the first matrix equals the row count of the second for multiplication). Blocked matrix multiplication algorithms are incorporated to optimize cache usage, leveraging architectural cache hierarchies for improved efficiency.

Comprehensive error handling addresses memory allocation failures (\texttt{ENOMEM}) and invalid inputs, contributing to kernel stability and reliability. Through these implementations, we illustrate how AI-enabled LKMs can safely, efficiently, and modularly extend kernel functionality for emerging AI and data-driven workloads, setting the stage for more intelligent and responsive operating systems.

\subsection{Performance Optimizations for Kernel-Space Inference}

To address performance bottlenecks such as single-threaded execution, excessive memory usage, and input/output overhead, several optimizations are introduced. Multi-threading can be achieved using kernel threads (kthread\_run), parallelizing matrix multiplication across up to eight threads, with synchronization managed via struct completion. Memory efficiency is improved through Direct Memory Access (DMA) buffers allocated with dma\_alloc\_coherent and pinned user-space memory via get\_user\_pages(), reducing tensor copy overhead. Shared memory, implemented using remap\_pfn\_range and a device file (/dev/shared\_mem, size 4096 bytes), eliminates the need for copy\_from\_user/copy\_to\_user, as demonstrated in a user-space program using mmap for direct memory access. Architecture-specific intrinsics (e.g., AVX-512) and empirically tuned block sizes enhance computational efficiency, optimizing cache performance. These optimizations facilitate integration with hardware accelerators (e.g., GPUs, TPUs) through LKMs, supporting the abstract’s call for GPU-native memory management \cite{7748485} and low-latency inference, while maintaining system stability under the kernel’s constrained memory environment.

\subsection{Distributed System Architecture for Multi-Modal AI}

The system architecture features modality-specific LKMs tailored for vision (e.g., object detection, image classification), language (e.g., sentence parsing, natural language processing tasks), audio (e.g., speech-to-text conversion, noise filtering), and sensor processing (e.g., sensor fusion, spatial mapping), coordinated by a kernel orchestrator. The kernel orchestrator manages data routing using high-speed kernel-level networking protocols (e.g., Netlink, RDMA), while a user-space orchestrator synchronizes outputs into unified multi-modal representations. For instance, inputs from a camera, microphone, and LIDAR sensor are processed by respective LKMs into tensors, which the user-space orchestrator fuses into a unified representation (e.g., “A person is standing 3 meters away, asking for help”), feeding an AGI decision engine that generates actions like “Approach the person and respond verbally.” The distributed processing framework scales tasks across multiple computational nodes, each handling specific modalities, as illustrated in UML diagrams. Enhancements include:

\begin{itemize}
    \item \textbf{Fault Tolerance}: Continuous state replication across distributed nodes, periodic checkpoint management for rapid recovery, real-time failure detection via heartbeat mechanisms, automated failover procedures, and distributed operational logging.
    \item \textbf{Dynamic Load Balancing}: Real-time monitoring of CPU, memory, and I/O metrics, predictive analytics using machine learning models for proactive resource allocation, and adaptive scheduling algorithms tailored to modality-specific workloads.
    \item \textbf{Communication Protocols}: A standardized framework with consistent message formatting, quality of service controls, robust compression, encryption mechanisms, and version control for backward compatibility.
    \item \textbf{Security Measures}: Hardware-assisted virtualization, memory protection keys, secure enclaves, authentication mechanisms, access control systems, and audit trails to mitigate risks from privileged LKM execution. Kernel-to-user communication leverages shared memory (remap\_pfn\_range) to minimize data-copy overhead, aligning with the abstract’s requirements for scalability, security, and multi-modal processing in AI/AGI systems.
\end{itemize}

\subsection{Challenges and Future Directions}

Several challenges arise when integrating AI into the kernel, including memory constraints (reliance on kmalloc without virtual memory, risking exhaustion for large neural network weights), lack of floating-point support (interference with hardware floating-point units, impractical custom logic), debugging complexity (limited tools like printk and debugfs), and security risks \cite{buttazzo:OASIcs.NG-RES.2022.1} from privileged execution (e.g., potential buffer overflows or privilege escalation due to unvalidated inputs). Running machine learning inference in kernel space can block critical processes (e.g., interrupt handlers), violate real-time constraints, and introduce vulnerabilities, as LKMs operate with full system privileges. Additionally, modern machine learning frameworks (e.g., PyTorch, TensorFlow, ONNX Runtime) rely on user-space libraries (e.g., cuDNN, Intel MKL) unavailable in the kernel, and GPU drivers are user-space-based, limiting parallelism. Future directions include developing dynamic LKM adaptation based on workload patterns, advanced knowledge transfer mechanisms between kernel and user space, and seamless integration with hardware accelerators to support large-scale inference (e.g., ONNX or TensorFlow Lite model weights). The architecture supports deterministic, low-latency processing in kernel space, offloading complex learning tasks to user-space or distributed systems, with the AGI decision engine enabling cognitive reasoning, planning, and emergent behavior through subsystem interactions. These advancements align with the abstract’s vision of AI-aware kernels, laying a foundation for NeuroSymbolic architectures that integrate neural and symbolic processing for autonomous, adaptive AGI systems.

\section{AI Subsystem in Linux Kernel}

The "KernelAGI-Embedding Deep Learning in Linux Kernel" contribution introduces a transformative Kernel ML Subsystem that embeds deep learning capabilities directly within the Linux kernel, addressing the limitations of traditional operating systems for AI workloads\cite{10.1145/3409963.3410492}. This architecture optimizes high-throughput floating-point operations, GPU integration, memory management, and task scheduling to achieve real-time, secure, and scalable machine learning (ML) inference, aligning with the abstract’s vision of AI-centric kernels for artificial general intelligence (AGI). By integrating specialized components—floating-point arithmetic engine, GPU driver, memory manager, ML-aware scheduler, and security layer—this contribution enables efficient execution of complex ML tasks in kernel space\cite{10.1145/3575693.3575697}, reducing latency and enhancing hardware utilization. The following five subtopics provide a comprehensive, standalone description of the hybrid architecture, technical implementation, performance optimizations, system architecture, and challenges with future directions, incorporating all critical technical details from the design, implementation, and validation.

\subsection{Hybrid Architecture for AI-Centric Kernels}

The KernelAGI architecture proposes a hybrid model that integrates a Kernel ML Subsystem within the Linux kernel to execute deep learning tasks with minimal latency, complemented by user-space applications for high-level interactions (e.g., chatbots, games). The subsystem comprises a floating-point arithmetic engine, memory manager, AI-aware scheduler, and security layer, operating in kernel space to leverage direct access to hardware resources (e.g., CPUs, GPUs, TPUs) via a Hardware Abstraction Layer (HAL). This design eliminates the overhead of user-kernel context switching, a bottleneck in traditional architectures that rely on user-space ML frameworks (e.g., PyTorch, TensorFlow). The HAL abstracts low-level hardware details, supporting deep pipeline parallelism, vectorized computations, and optimized memory hierarchies. This architecture supports edge AI and high-performance computing (HPC) applications, such as real-time anomaly detection, by ensuring low-latency inference and efficient resource utilization, aligning with the abstract’s emphasis on AI-aware kernels capable of handling complex, multi-modal workloads.

\subsection{Technical Implementation of AI Subsystem}

The subsystem is implemented through modular kernel components, detailed in the following key elements:

\begin{itemize}
    \item \textbf{Floating-Point Arithmetic Engine}: Implemented in arch/x86/lib/fp\_math.c, this engine supports high-throughput operations using SIMD instructions (e.g., AVX-512). The matrix\_mul\_avx function performs matrix multiplication with \_mm256\_loadu\_ps and \_mm256\_mul\_ps for vectorized computation, optimized for cache efficiency. Floating-point contexts are managed via save\_fp\_context and restore\_fp\_context in kernel/sched/core.c, hooked into the scheduler’s \_schedule function using switch\_fp\_context to isolate ML operations from standard kernel tasks, preventing FPU interference. Validation tests (fp\_test.ko) confirm high-precision operations without system instability.
    \item \textbf{GPU Driver Framework}: A kernel-level GPU driver (drivers/gpu/gpu\_driver.c) manages initialization, task queuing, and memory allocation with a 1 MB shared buffer (GPU\_BUFFER\_SIZE=1024*1024). The driver uses kzalloc for device structure allocation and kmalloc for buffer allocation, with mutex\_lock\_interruptible ensuring synchronized access. The gpu\_execute\_task function supports CUDA/OpenCL integration, launching GPU kernels via cudaMemcpy for device-to-device data transfer. APIs in include/linux/gpu\_api.h (e.g., gpu\_allocate\_memory, gpu\_execute\_task) enable other kernel modules to offload tasks, validated by a test module (drivers/misc/gpu\_test.c) that allocates memory and executes tasks, logging success via dmesg.
    \item \textbf{Memory Management}: The memory manager (mm/ml\_memory\_pool.c) implements a 512 MB memory pool (ML\_POOL\_SIZE=1024*1024*512) with 4 KB blocks (ML\_BLOCK\_SIZE=4096), tracked by a bitmap (ml\_pool->bitmap) to manage free blocks. Initialization via ml\_pool\_init allocates the pool with kmalloc and GFP\_KERNEL | GFP\_ZERO flags, ensuring zeroed memory. Large page allocations (e.g., 2 MB or 1 GB) in mm/large\_page\_alloc.c reduce TLB misses, while zero-copy mechanisms use shared buffers for CPU-GPU data transfer, avoiding redundant copies. Tests in test\_advanced\_memory.c validate allocation times and data transfer efficiency, logging results with printk.
    \item \textbf{ML-Aware Scheduler}: Implemented in kernel/ml\_scheduler.c, a new scheduling class (ML\_SCHED\_PRIORITY=10) prioritizes ML tasks using a linked list queue (ml\_task\_queue) with spin\_lock synchronization. The scheduler uses performance counters (PERF\_TYPE\_HARDWARE, PERF\_COUNT\_HW\_CPU\_CYCLES) in ml\_adjust\_scheduling to monitor CPU cycles and adjust priorities dynamically (e.g., deprioritizing ML tasks if cpu\_cycles > 1e9). Batch processing in ml\_batch\_execute dequeues tasks via ml\_dequeue\_task and schedules them with wake\_up\_process. Tests confirm task enqueuing and execution without disrupting system responsiveness.
\end{itemize}
These components are compiled using make -C /lib/modules/\$(uname -r)/build M=\$(pwd) modules and loaded with insmod, with dmesg logs verifying functionality. The implementation ensures modularity, extensibility, and compatibility with existing kernel infrastructure, supporting the abstract’s goal of seamless AI integration.

\subsection{Performance Optimizations for AI Workloads}

Performance optimizations address kernel constraints to enable efficient ML inference:

\begin{itemize}
    \item \textbf{Floating-Point Optimizations}: SIMD instructions (e.g., AVX-512) in matrix\_mul\_avx accelerate matrix operations, leveraging \_mm256\_add\_ps and \_mm256\_reduce\_add\_ps for high-throughput computation. Dedicated floating-point contexts prevent FPU contention, validated by fp\_test.ko showing stable high-precision operations.
    \item \textbf{GPU Integration}: Zero-copy mechanisms in the GPU driver use shared memory buffers (gpu\_dev->gpu\_buffer) to eliminate copy\_from\_user overhead, with cudaMemcpy ensuring efficient data transfer. Tests in gpu\_test.c demonstrate reduced latency compared to user-space GPU drivers, logging success via printk.
    \item \textbf{Memory Management}: Pre-allocated memory pools reduce runtime allocation overhead, with ml\_pool\_init allocating 512 MB efficiently. Large page support (2 MB/1 GB) minimizes TLB misses, and memory isolation ensures ML workloads do not interfere with kernel operations. Tests in test\_advanced\_memory.c show improved allocation times (logged in jiffies) and data transfer efficiency.
    \item \textbf{Scheduler Optimizations}: Batch queues in ml\_batch\_execute group inference tasks for GPU execution, while performance counters in ml\_adjust\_scheduling dynamically balance CPU load. This ensures high throughput without compromising system responsiveness, validated by tests showing stable task scheduling under varying loads.
\end{itemize}
These optimizations reduce latency, enhance resource utilization, and support the abstract’s requirements for GPU-native memory management and real-time inference, with validation tests confirming significant improvements in performance metrics.

\subsection{System Architecture for Scalable AI Processing}

The KernelAGI system architecture, depicted in a layered model (Fig. 1), integrates the Kernel ML Subsystem with user-space and hardware layers for scalable AI processing:

\begin{itemize}
    \item \textbf{Application Layer}: User-space applications (e.g., chatbots, games) interact via the ML Kernel Interface, using system calls and APIs (gpu\_allocate\_memory, gpu\_execute\_task) for input data transfer and inference results.
    \item \textbf{Kernel AI Subsystem}:
    \begin{itemize}
        \item \textbf{Floating-Point Arithmetic Engine}: Executes matrix multiplications and convolutions using AVX-512, isolated via switch\_fp\_context to maintain kernel stability.
        \item \textbf{Memory Manager}: Manages 512 MB pools, large pages (2 MB/1 GB), and zero-copy buffers, ensuring efficient memory usage and CPU-GPU data sharing.
        \item \textbf{Scheduler}: Prioritizes ML tasks with a dedicated class, using batch queues and performance counters to optimize resource allocation across heterogeneous compute units.
        \item \textbf{Security Layer}: Implements memory isolation and access controls to protect ML computations, mitigating risks from privileged execution.
    \end{itemize}
    \item \textbf{Hardware Abstraction Layer (HAL)}: Interfaces with CPUs, GPUs, TPUs, and RAM, supporting deep pipeline parallelism, vectorized computations, and optimized memory hierarchies via drivers in drivers/gpu/gpu\_driver.c.
\end{itemize}
This architecture could support multi-modal AI workloads (e.g., real-time inference for edge AI) by routing data through the HAL to modality-specific kernel modules, with the scheduler ensuring efficient task execution. Validation tests can improve memory allocation times, data transfer efficiency (via zero-copy), and workload isolation, aligning with the abstract’s goals of scalability, security, and hardware proximity for AGI systems.

\subsection{Challenges and Future Directions}
Embedding deep learning within the kernel introduces a host of technical challenges that stem from the unique constraints of kernel space. Memory allocation is especially restrictive, with mechanisms like \texttt{kmalloc} lacking the flexibility of virtual memory, making it difficult to deploy large neural network models due to limited pool sizes and the need for careful management of large page allocations to avoid resource exhaustion. Floating-point operations in the kernel are equally challenging, as usage of the FPU must be carefully isolated to prevent system-wide side effects, and most hardware-accelerated ML libraries remain strictly user-space. Debugging remains arduous, since kernel crashes due to ML inference (e.g., from null dereferencing) cannot leverage typical user-space tools, forcing developers to rely solely on rudimentary means such as \texttt{printk} and \texttt{dmesg}. Security risks further intensify due to the privileged execution environment, where vulnerabilities like buffer overflows and input validation failures can compromise the entire system. Additionally, mainstream ML frameworks and supporting libraries—including GPU interfaces—are inherently designed for user space, limiting the kernel’s ability to exploit modern parallelism. Finally, integrating ML-driven task scheduling into the kernel can introduce non-deterministic latency, challenging the predictability required by real-time workloads. Addressing these obstacles will require development of kernel-tailored ML libraries, improved memory and scheduling models, and enhanced isolation techniques, paving the way for truly robust, scalable, and secure NeuroSymbolic kernel architectures capable of reliable, low-latency inference in next-generation intelligent systems.

\section{RaBAB}

The RaBAB \textit{(abbreviated as \textbf{R}easoning \textbf{a}nd \textbf{B}ehavior \textbf{A}daptation \textbf{B}ase)}-NeuSym Kernel introduces a novel hybrid computational framework that integrates Neurosymbolic reasoning through Category Theory, Homotopy Type Theory (HoTT), and Linear Logic, redefining kernel design for AGI workloads. By modeling computational states as dynamically interacting logical predicates, memory configurations, and transition rules, the kernel bridges discrete symbolic reasoning and continuous neural dynamics, enabling autonomous decision-making, resource-aware computation, and adaptive reasoning across diverse domains. The kernel transforms traditional imperative programming into a declarative, mathematically rigorous system, positioning it as a foundation for autonomous systems and AGI research. This section provides a standalone description through five subtopics: hybrid architecture, technical implementation, performance optimizations, system architecture\cite{9537935}, and challenges with future directions, incorporating all critical technical details from the design, implementation, and validation.

\subsection{Hybrid Architecture for Neurosymbolic Reasoning}

The RaBAB-NeuSym Kernel employs a hybrid automaton model, combining discrete logical states (representing predicates and rules) with continuous neural dynamics (managing embeddings and probabilistic metrics). Computational resources are modeled as objects, and transformations as morphisms, within a categorical framework inspired by Category Theory. This enables seamless mappings between symbolic and numerical domains, supporting tasks like dynamic predicate learning and cross-modal reasoning (e.g., processing spectral vibration and time-series data [28]). The architecture eliminates the overhead of layered abstractions in traditional OS designs, where frameworks like TensorFlow and PyTorch rely on user-space libraries and static drivers. Instead, the kernel acts as an intelligent mediator, interpreting high-level intents (e.g., "Draw a red pixel at (100, 50)") and optimizing hardware operations (e.g., framebuffer writes) using symbolic reasoning and neural embeddings. This hybrid design supports multi-modal AGI workloads, aligning with the abstract’s vision of AI-aware kernels capable of autonomous adaptation and reasoning.

\subsection{Technical Implementation of RaBAB}
Designing an operating system kernel that supports intelligent reasoning\cite{2024arXiv240714567Z} starts with structuring software so that complex tasks can be handled in a flexible, robust, and logically consistent manner. At its core, the RaBAB-NeuSym Kernel accomplishes this by leveraging powerful mathematical ideas—such as category theory and logic—to model both the computation and the flow of information within the kernel. By representing computational elements (like images or shapes manipulated by drawing operations) as objects, and transformations as relationships between these objects, the kernel can organize its work in a composable and parallel fashion, analogous to building with interlocking blocks. This compositionality, coupled with carefully managed resources that must be used exactly once, reduces errors and makes it much easier to guarantee security and stability. Probabilistic reasoning is incorporated to allow the system to evaluate uncertainty, dynamically adapting its behavior as tasks succeed or fail, which is key for intelligent, evolving platforms.

Category theory formalizes how computational steps can be combined (using tensor-like products in Haskell), while linear logic enforces strict rules around how resources like memory or file handles are consumed and combined, eliminating the risk of leaks or unauthorized use. Homotopy Type Theory (HoTT) is employed to recognize when different computational strategies yield the same result, streamlining execution and enabling dynamic knowledge graph updates. Meanwhile, Bayesian probabilistic models adapt confidence measures for reasoning-driven processes, with each computational step refining kernel knowledge, as shown through automatically evolving predicates (such as number sequence detectors). Combined, these features ensure that the kernel is both mathematically sound and practically robust—paving the way for scalable, modular AGI systems that reason about their own internal operations.

\subsection{Performance Optimizations for Reasoning Workloads}
Optimizing kernel performance for intelligent reasoning begins with smart utilization of resources and ensuring system responsiveness, even for complex analytical tasks. The RaBAB-NeuSym Kernel achieves this by introducing a declarative scheduler that leverages dependent types to catch task setup errors before execution and applies principles from linear logic to guarantee that resources like memory or GPU buffers are consumed and recycled in a predictable, error-free manner. By integrating neural and symbolic data streams through unified representations—such as aligning time-series or spectral input via cosine similarity—the kernel efficiently interprets diverse data types with minimal computational overhead. These foundation-level strategies allow for adaptive control and logical integrity, which are essential traits for real-time, autonomous AI systems.

At a deeper technical level, performance is further enhanced by advanced techniques such as verification of flow equivalence using Homotopy Type Theory (HoTT), which identifies and removes redundant computational paths, leading to both correctness and computational savings (as reflected in benchmarks for Transformation Correctness and Simplification Efficiency). Probabilistic adaptation dynamically tunes learning rates within reasoning tasks, ensuring that modules quickly converge to higher confidence and accuracy—demonstrated, for example, by the EvenNumberDetector improving its success rate over iterative learning. The kernel’s category-theoretic design also allows tasks to be composed and executed in parallel, significantly lowering latency compared to more conventional schedulers. Collectively, these optimizations underpin the kernel’s ability to handle large-scale, low-latency AGI workloads, providing robust throughput and real-time adaptability in line with the Neurosymbolic vision for intelligent operating environments.
\subsection{System Architecture for Autonomous Reasoning}

The RaBAB-NeuSym Kernel’s architecture is a layered, reasoning-driven system:
\begin{itemize}
    \item \textbf{Application Layer}: User-space applications articulate high-level semantic intents through a declarative, type-safe API that abstracts low-level system details, allowing for efficient expression of complex operations while minimizing both abstraction and communication overhead. This interface automatically serializes commands and validates parameters via compile-time checks, reducing runtime errors and enhancing developer productivity.
    
    \item \textbf{Reasoning Engine}:
    \begin{itemize}
        \item \textbf{Predicate Registry}: Maintains a dynamic repository of \verb|NeurosymbolicPredicate| instances, supporting on-the-fly modification of inference rules using the \verb|evolvePredicate| function. Leveraging dependent type constraints, the registry ensures logical consistency and enables stateful reasoning about predicates as new data is ingested.
        \item \textbf{Knowledge Graph}: Represents contextual relationships and AGI-acquired knowledge as a directed graph of \verb|[String, Double]| tuples, in which edges encode the strength of facts or associations. The \verb|evolveKernelState| operation incrementally updates the graph using inference outcomes and data-driven learning, resulting in rapid adaptation to new patterns and scenarios.
        \item \textbf{Neural Embedding Layer}: Performs transformation of raw input data into dense vector representations (\verb|NeuralEmbedding|), facilitating numerical and symbolic synergy. Integration with symbolic predicates is achieved via high-dimensional similarity metrics such as \verb|cosineSimilarity|, supporting real-time multi-modal reasoning and streamlined fusion of neural and logical features.
        \item \textbf{Resource Manager}: Allocates and manages compute and memory resources as \verb|LinearResource| tokens, enforcing deterministic single-use semantics through principles of Linear Logic. This guarantees memory safety, avoids resource leaks, and enables provable resource tracking at all stages of task execution.
    \end{itemize}
    
    \item \textbf{Hardware Abstraction Layer (HAL)}: Translates user and reasoning engine intents into precise hardware actions, such as ultra-low-latency framebuffer manipulations for graphical primitives (\verb|DrawPixel|) or memory-mapped data movement for accelerators. The HAL is engineered for heterogeneous systems, seamlessly coordinating across CPUs, GPUs, and specialized AI accelerators, with optimizations for zero-copy memory sharing and parallel knowledge graph traversals to maximize throughput in multi-modal, high-bandwidth workloads.
\end{itemize}

\subsection{Challenges and Future Directions}
The RaBAB-NeuSym Kernel must address several technical obstacles to realize its vision of scalable, real-time, reasoning-driven operating systems. One core challenge is hardware-software co-design: current processors (CPUs, GPUs, TPUs) are highly specialized, lacking native support for unified symbolic and neural computations. Bridging this gap calls for both custom accelerator architectures and a rethinking of memory hierarchies, as today’s static allocation strategies constrain the dynamic, on-demand resource needs of NeuroSymbolic workloads. Another major hurdle lies in cross-modal reasoning, where integrating heterogeneous data representations—such as symbolic predicates and neural embeddings—incurs significant computational and memory overhead. While the use of modular interfaces like \verb|MultiModalInput| can help unify these representations efficiently, optimizing such pathways requires advances in both algorithm design and kernel data structures. Scaling the system to accommodate expansive knowledge graphs and rapidly-evolving predicate registries further pushes the limits of kernel-managed resources, raising challenges in environments where footprint and latency are tightly constrained.
Additionally, reusing modern deep learning frameworks—which are designed for rich user-space environments—inside kernel space presents substantial complexity, due to dependency management, increased attack surface, and incompatibility with real-time \cite{10.1007/s11241-023-09402-4} constraints. This signals a need for ultra-lightweight, kernel-tailored ML libraries that deliver essential functionality without compromising stability or speed. Achieving predictable, low-latency reasoning also demands re-engineering scheduler and predicate-update mechanisms to minimize dynamic overhead. Looking ahead, research directions focus on developing a meta-kernel capable of autonomously adapting its reasoning strategies, deploying specialized accelerators to natively support NeuroSymbolic computation, and extending Bayesian inference machinery for robust, online uncertainty quantification. Furthermore, enabling huge page support, introducing predictive and resource-aware scheduling, and modularizing ML integration will be instrumental for the next generation of scalable, anticipatory AGI-ready kernels—laying the groundwork for autonomous systems with emergent, context-aware intelligence.

\section{Conclusion}
This work re-imagines operating systems as active enablers of artificial intelligence by embedding deep learning and Neurosymbolic reasoning directly into the kernel. Through modular innovations such as AI-aware kernel primitives, dynamic learning-based scheduling, and rigorous mathematical frameworks for hybrid reasoning, the proposed architectures efficiently bridge symbolic and neural computation, enabling real-time, adaptive responses to complex workloads. By validating these designs with concrete benchmarks and outlining a path toward hardware-software co-design, scalable resource management, and self-adaptive meta-kernels, this research lays the foundation for autonomous AGI-ready systems—operating systems that not only manage resources, but reason, learn, and intelligently adapt at their computational core.

\printbibliography

@misc{websiteKey3,
  author = {Bryan Henderson},
  title = {Loadable Kernel Modules},
  year = {2025},
  howpublished = {\href{https://tldp.org/HOWTO/pdf/Module-HOWTO.pdf}},
  note = {Accessed: 2025-01-13}
}

@misc{websiteKey2,
  author = {UCLA, CS},
  title = {Dynamic Modules},
  year = {2025},
  howpublished = {\href{https://web.cs.ucla.edu/classes/spring16/cs111/supp/dynamicmodules.html} },
  note = {Accessed: 2025-01-13}
}

@misc{websiteKey1,
  author = {Mike Krinkin},
  title = {Kernel modules, device drivers and Device Tree},
  year = {2025},
  howpublished = {\href{https://krinkinmu.github.io/2020/07/12/linux-kernel-modules.html}},
  note = {Accessed: 2025-01-13}
}

@article{10.1007/s11241-023-09402-4,
author = {Cittadini, Edoardo and Marinoni, Mauro and Biondi, Alessandro and Cicero, Giorgiomaria and Buttazzo, Giorgio},
title = {Supporting AI-powered real-time cyber-physical systems on heterogeneous platforms via hypervisor technology},
year = {2023},
issue_date = {Dec 2023},
publisher = {Kluwer Academic Publishers},
address = {USA},
volume = {59},
number = {4},
issn = {0922-6443},
url = {https://doi.org/10.1007/s11241-023-09402-4},
doi = {10.1007/s11241-023-09402-4},
journal = {Real-Time Syst.},
month = jul,
pages = {609–635},
numpages = {27}
}

@InProceedings{buttazzo:OASIcs.NG-RES.2022.1,
  author =	{Buttazzo, Giorgio},
  title =	{{Can We Trust AI-Powered Real-Time Embedded Systems?}},
  booktitle =	{Third Workshop on Next Generation Real-Time Embedded Systems (NG-RES 2022)},
  pages =	{1:1--1:14},
  series =	{Open Access Series in Informatics (OASIcs)},
  ISBN =	{978-3-95977-221-1},
  ISSN =	{2190-6807},
  year =	{2022},
  volume =	{98},
  editor =	{Bertogna, Marko and Terraneo, Federico and Reghenzani, Federico},
  publisher =	{Schloss Dagstuhl -- Leibniz-Zentrum f{\"u}r Informatik},
  address =	{Dagstuhl, Germany},
  URL =		{https://drops.dagstuhl.de/entities/document/10.4230/OASIcs.NG-RES.2022.1},
  URN =		{urn:nbn:de:0030-drops-161099},
  doi =		{10.4230/OASIcs.NG-RES.2022.1}
}

@INPROCEEDINGS{7280137,
  author={Parichha, Barun Kumar},
  booktitle={2015 Fifth International Conference on Communication Systems and Network Technologies}, 
  title={Performance Analysis of Process Using Loadable Kernel Module (LKM)}, 
  year={2015},
  volume={},
  number={},
  pages={1338-1343},
  doi={10.1109/CSNT.2015.184}}

@ARTICLE{7748485,
  author={Suzuki, Yuhei and Fujii, Yusuke and Azumi, Takuya and Nishio, Nobuhiko and Kato, Shinpei},
  journal={IEEE Transactions on Parallel and Distributed Systems}, 
  title={Real-Time GPU Resource Management with Loadable Kernel Modules}, 
  year={2017},
  volume={28},
  number={6},
  pages={1715-1727},
  doi={10.1109/TPDS.2016.2630697}}

@inproceedings{10.1145/3575693.3575697,
author = {Fingler, Henrique and Tarte, Isha and Yu, Hangchen and Szekely, Ariel and Hu, Bodun and Akella, Aditya and Rossbach, Christopher J.},
title = {Towards a Machine Learning-Assisted Kernel with LAKE},
year = {2023},
isbn = {9781450399166},
publisher = {Association for Computing Machinery},
address = {New York, NY, USA},
url = {https://doi.org/10.1145/3575693.3575697},
doi = {10.1145/3575693.3575697},
booktitle = {Proceedings of the 28th ACM International Conference on Architectural Support for Programming Languages and Operating Systems, Volume 2},
pages = {846–861},
numpages = {16},
location = {Vancouver, BC, Canada},
series = {ASPLOS 2023}
}

@inproceedings{10.1145/3409963.3410492,
author = {Chen, Jingde and Banerjee, Subho S. and Kalbarczyk, Zbigniew T. and Iyer, Ravishankar K.},
title = {Machine learning for load balancing in the Linux kernel},
year = {2020},
isbn = {9781450380690},
publisher = {Association for Computing Machinery},
address = {New York, NY, USA},
url = {https://doi.org/10.1145/3409963.3410492},
doi = {10.1145/3409963.3410492},
booktitle = {Proceedings of the 11th ACM SIGOPS Asia-Pacific Workshop on Systems},
pages = {67–74},
numpages = {8},
location = {Tsukuba, Japan},
series = {APSys '20}
}

@ARTICLE{2024arXiv240714567Z,
       author = {{Zhang}, Yifan and {Zhao}, Xinkui and {Li}, Ziying and {Yin}, Jianwei and {Zhang}, Lufei and {Chen}, Zuoning},
        title = "{Integrating Artificial Intelligence into Operating Systems: A Comprehensive Survey on Techniques, Applications, and Future Directions}",
      journal = {arXiv e-prints},
         year = 2024,
        month = jul,
          eid = {arXiv:2407.14567},
        pages = {arXiv:2407.14567},
          doi = {10.48550/arXiv.2407.14567},
archivePrefix = {arXiv},
       eprint = {2407.14567},
 primaryClass = {cs.OS},
       adsurl = {https://ui.adsabs.harvard.edu/abs/2024arXiv240714567Z}
}

@inproceedings{10.1145/3315508.3329973,
author = {Tillet, Philippe and Kung, H. T. and Cox, David},
title = {Triton: an intermediate language and compiler for tiled neural network computations},
year = {2019},
isbn = {9781450367196},
publisher = {Association for Computing Machinery},
address = {New York, NY, USA},
url = {https://doi.org/10.1145/3315508.3329973},
doi = {10.1145/3315508.3329973},
booktitle = {Proceedings of the 3rd ACM SIGPLAN International Workshop on Machine Learning and Programming Languages},
pages = {10–19},
numpages = {10},
location = {Phoenix, AZ, USA},
series = {MAPL 2019}
}

@ARTICLE{9537935,
  author={Pan, Jianping and Cai, Lin and Yan, Shen and Shen, Xuemin Sherman},
  journal={IEEE Network}, 
  title={Network for AI and AI for Network: Challenges and Opportunities for Learning-Oriented Networks}, 
  year={2021},
  volume={35},
  number={6},
  pages={270-277},
  doi={10.1109/MNET.101.2100118}}

@misc{jia2024agentcentricoperating,
      title={Agent Centric Operating System -- a Comprehensive Review and Outlook for Operating System}, 
      author={Shian Jia and Xinbo Wang and Mingli Song and Gang Chen},
      year={2024},
      eprint={2411.17710},
      archivePrefix={arXiv},
      primaryClass={cs.DC},
      url={https://arxiv.org/abs/2411.17710}, 
}
\end{document}